%% file: paper.tex
\documentclass[format=acmsmall,timestamp=false,screen]{acmart}
\pdfoutput=1

\newif\ifarxiv
\arxivtrue

\ifarxiv
\usepackage{microtype}
\setkeys{acmart.cls}{nonacm}
\else
\setkeys{acmart.cls}{draft}
\fi
\setcopyright{none}

\acmDOI{}

\usepackage[utf8]{inputenc}
\usepackage{array}
\usepackage[lined,boxed,commentsnumbered]{algorithm2e}
\usepackage{amsmath}
\usepackage[english]{babel}
\usepackage{enumitem}
\usepackage{hyperref}
\usepackage{cleveref}
\usepackage{verbatim}
\usepackage{subcaption}
\usepackage{booktabs}
\usepackage{xspace}

\ifarxiv
\usepackage[frozencache]{minted}
\else
\usepackage[finalizecache]{minted}
\fi
\usepackage{mathtools}

\usepackage{tikz}
\usetikzlibrary{cd}
\usetikzlibrary{calc, positioning, trees, plotmarks}
\usetikzlibrary{shapes, shapes.geometric, decorations.pathreplacing, decorations.markings}
\usetikzlibrary{arrows, arrows.meta, chains, fit, backgrounds, scopes}
\newcommand{\convexpath}[2]{
  [
  create hullcoords/.code={
    \global\edef\namelist{#1}
    \foreach [count=\counter] \nodename in \namelist {
      \global\edef\numberofnodes{\counter}
      \coordinate (hullcoord\counter) at (\nodename);
    }
    \coordinate (hullcoord0) at (hullcoord\numberofnodes);
    \pgfmathtruncatemacro\lastnumber{\numberofnodes+1}
    \coordinate (hullcoord\lastnumber) at (hullcoord1);
  },
  create hullcoords
  ]
  ($(hullcoord1)!#2!-90:(hullcoord0)$)
  \foreach [
  evaluate=\currentnode as \previousnode using \currentnode-1,
  evaluate=\currentnode as \nextnode using \currentnode+1
  ] \currentnode in {1,...,\numberofnodes} {
    let \p1 = ($(hullcoord\currentnode) - (hullcoord\previousnode)$),
    \n1 = {atan2(\y1,\x1) + 90},
    \p2 = ($(hullcoord\nextnode) - (hullcoord\currentnode)$),
    \n2 = {atan2(\y2,\x2) + 90},
    \n{delta} = {Mod(\n2-\n1,360) - 360}
    in
    {arc [start angle=\n1, delta angle=\n{delta}, radius=#2]}
    -- ($(hullcoord\nextnode)!#2!-90:(hullcoord\currentnode)$)
  }
}

\newcommand{\ltwo}{\ensuremath{{L}^2(\Omega)}\xspace}
\newcommand{\ltwozero}{\ensuremath{{L}^2_0(\Omega)}\xspace}

\newcommand{\htwo}{\ensuremath{{H}^2(\Omega)}\xspace}
\newcommand{\honev}{\ensuremath{{H}^1(\Omega; \mathbb{R}^d)}\xspace}
\newcommand{\honezerov}{\ensuremath{{H}^1_0(\Omega; \mathbb{R}^d)}\xspace}

\newcommand{\honehalfDv}{\ensuremath{{H}^{1/2}(\partial \Omega_D; \mathbb{R}^d)}\xspace}
\newcommand{\hdiv}{\ensuremath{{H}(\mathrm{div}, \Omega)}\xspace}
\newcommand{\hcurl}{\ensuremath{{H}(\mathrm{curl}, \Omega)}\xspace}

\definecolor{blue}{rgb}{0.2980392156862745, 0.4470588235294118, 0.6901960784313725}
\definecolor{green}{rgb}{0.3333333333333333, 0.6588235294117647, 0.40784313725490196}
\definecolor{red}{rgb}{0.7686274509803922, 0.3058823529411765, 0.3215686274509804}
\definecolor{purple}{rgb}{0.5058823529411764, 0.4470588235294118, 0.6980392156862745}
\definecolor{yellow}{rgb}{0.8, 0.7254901960784313, 0.4549019607843137}
\definecolor{lightblue}{rgb}{0.39215686274509803, 0.7098039215686275, 0.803921568627451}
\definecolor{orange}{rgb}{0.8666666666666667, 0.5176470588235295, 0.3215686274509804}
\definecolor{brown}{rgb}{0.5764705882352941, 0.47058823529411764, 0.3764705882352941}
\definecolor{pink}{rgb}{0.8549019607843137, 0.5450980392156862, 0.7647058823529411}
\definecolor{gray}{rgb}{0.5490196078431373, 0.5490196078431373, 0.5490196078431373}

\let\div\relax
\DeclareMathOperator{\div}{div}
\DeclareMathOperator{\curl}{curl}
\DeclareMathOperator{\grad}{grad}

\acmJournal{TOMS}

\newcommand{\pcpatch}{\texttt{PCPATCH}\xspace}
\newcommand{\dmplex}{\texttt{DMPLEX}\xspace}
\newcommand{\snespatch}{\texttt{SNESPATCH}\xspace}

\title{\pcpatch: software for the topological construction of multigrid relaxation methods}

\author{Patrick E.~Farrell}
\orcid{0000-0002-1241-7060}
\affiliation{%
  \institution{University of Oxford}
  \department{Mathematical Institute}}
\email{patrick.farrell@maths.ox.ac.uk}

\author{Matthew G.~Knepley}
\orcid{0000-0002-2292-0735}
\affiliation{%
  \institution{University at Buffalo}
  \department{Department of Computer Science and Engineering}}
\email{knepley@gmail.com}

\author{Lawrence Mitchell}
\orcid{0000-0001-8062-1453}
\affiliation{%
  \institution{Durham University}
  \department{Department of Computer Science}}
\email{lawrence.mitchell@durham.ac.uk}

\author{Florian Wechsung}
\orcid{0000-0003-2195-6522}
\affiliation{%
  \institution{New York University}
  \department{Courant Institute of Mathematical Sciences}}
\email{wechsung@cims.nyu.edu}

\numberwithin{equation}{section}

\crefname{algorithm}{algorithm}{algorithms}
\crefname{figure}{figure}{figures}
\crefname{table}{table}{tables}

\begin{abstract}
  Effective relaxation methods are necessary for good multigrid convergence.
  For many equations, standard Jacobi and Gau\ss--Seidel are inadequate, and
  more sophisticated space decompositions are required; examples include
  problems with semidefinite terms or saddle point structure. In this
  paper we present a unifying software abstraction, \pcpatch, for the topological
  construction of space decompositions for multigrid relaxation methods.
  Space decompositions are specified by collecting topological
  entities in a mesh (such as all vertices or faces) and applying a construction rule (such
  as taking all degrees of freedom in the cells around each entity). The
  software is implemented in PETSc and facilitates the elegant expression of a
  wide range of schemes merely by varying solver options at runtime. In turn,
  this allows for the very rapid development of fast solvers for difficult problems.
\end{abstract}

\ifarxiv
\else
\fi


\begin{document}
\maketitle

\section{Introduction}\label{sec:introduction}

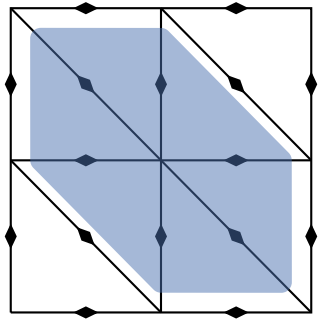
\begin{figure}[htbp]
  \centering
  \begin{tikzpicture}[scale=4,
    on each segment/.style={
      decorate,
      decoration={
        show path construction,
        moveto code={},
        lineto code={
          \path [#1]
          (\tikzinputsegmentfirst) -- (\tikzinputsegmentlast);
        }}},
    nedof/.style={postaction={decorate, decoration={markings,
          mark=at position 0.5 with {
            \node[transform shape,shape=diamond,fill,draw, inner
            sep=0pt, minimum height=4pt, minimum width=8pt] {};
          }}}}]
    \foreach \i/\k in {0/0, 0.5/1, 1/2} {
      \foreach \j/\l in {0/0, 0.5/1, 1/2} {
        \coordinate (v\k\l) at (\i, \j);
      }
    }

    \draw[thick, black, line join=miter, postaction={on each segment={nedof}}] (v00)
    -- (v10) -- (v20) -- (v21) -- (v22) -- (v12) -- (v02) -- (v01)
    -- (v00);
    \draw[nedof, thick, black, line join=cap] (v01) -- (v11);
    \draw[nedof, thick, black, line join=cap] (v11) -- (v21);
    \draw[nedof, thick, black, line join=cap] (v10) -- (v11);
    \draw[nedof, thick, black, line join=cap] (v11) -- (v12);
    \draw[nedof, thick, black, line join=cap] (v01) -- (v10);
    \draw[nedof, thick, black, line join=cap] (v02) -- (v11);
    \draw[nedof, thick, black, line join=cap] (v11) -- (v20);
    \draw[nedof, thick, black, line join=cap] (v12) -- (v21);
    \coordinate (i10) at ($(v10) + (90:0.1)$);
    \coordinate (i20) at ($(v20) + (135:0.141)$);
    \coordinate (i21) at ($(v21) + (180:0.1)$);
    \coordinate (i12) at ($(v12) + (270:0.1)$);
    \coordinate (i02) at ($(v02) + (315:0.141)$);
    \coordinate (i01) at ($(v01) + (0:0.1)$);
    \path[fill=blue, opacity=0.5]
    \convexpath{i01,i02,i12,i21,i20,i10}{1pt};
  \end{tikzpicture}
  \caption{Patch of cells resulting in robust relaxation for the
    problem of~\eqref{eq:hcurl}.\label{fig:hcurl-patch}}
\end{figure}

It is well known that geometric multigrid with Jacobi relaxation
is an effective solver for many problems. For example, when applied to
discretisations of: given a Lipschitz domain $\Omega \subset \mathbb{R}^3$, $f \in L^2(\Omega; \mathbb{R}^3)$ and $\alpha > 0$, find $u \in H^1(\Omega; \mathbb{R}^3)$ such that
\begin{equation}
  \label{eq:hgrad}
(u, v) + \alpha (\nabla u, \nabla v) = (f, v) \text{ for all } v \in H^1(\Omega; \mathbb{R}^3),
\end{equation}
geometric multigrid with Jacobi relaxation gives mesh-independent and $\alpha$-robust
convergence.
However, for many other problems, this is not the case. For example,
when applied to discretisations of: find $u \in \hcurl$ such that
\begin{equation}
  \label{eq:hcurl}
(u, v) + \alpha (\nabla \times u, \nabla \times v) = (f, v) \text{ for all } v \in \hcurl,
\end{equation}
geometric multigrid with Jacobi relaxation gives neither mesh-independent
nor $\alpha$-robust convergence. A simple modification restores both properties:
use a relaxation method that solves simultaneously for
all degrees of freedom in the patch of cells around each vertex, excluding those
on the boundary of the patch \citep{arnold2000}. The patch for the lowest-order
N\'ed\'elec element of the first kind is shown in \cref{fig:hcurl-patch}.

This same relaxation, which we refer to as a \emph{vertex-star} iteration, arises
in other contexts. It yields mesh-independent and parameter-robust convergence
for the \hdiv and \hcurl Riesz maps \citep{arnold1997,arnold2000}, for nearly incompressible linear elasticity
and Reissner--Mindlin plates
\citep{schoberl1999,schoberl1999b}, and for the Navier--Stokes equations \citep{benzi2006,farrell2018b}.
When combined with a low-order solver, it yields $p$-independent convergence for
high-order discretisations of symmetric second-order elliptic problems \citep{pavarino1993}.

\begin{figure}[htbp]
  \centering
  \begin{tikzpicture}[scale=4]
    \foreach \i/\k in {0/0, 0.5/1, 1/2} {
      \foreach \j/\l in {0/0, 0.5/1, 1/2} {
        \coordinate (v\k\l) at (\i, \j);
      }
    }
    \coordinate (o20) at ($(v20) + (315:0.1)$);
    \coordinate (o21) at ($(v21) + (0:0.0707)$);
    \coordinate (o12) at ($(v12) + (90:0.0707)$);
    \coordinate (o02) at ($(v02) + (135:0.1)$);
    \coordinate (o01) at ($(v01) + (180:0.0707)$);
    \coordinate (o10) at ($(v10) + (270:0.0707)$);

    \path[fill=blue, opacity=0.5]
    \convexpath{o10,o01,o02,o12,o21,o20}{1pt};


    \foreach \i/\k in {0/0, 0.5/1, 1/2} {
      \foreach \j/\l in {0/0, 0.5/1, 1/2} {
        \node[shape=diamond,draw,green, fill=green,inner
        sep=0pt,
        minimum height=14pt, minimum width=14pt] at (v\k\l) {};
        \draw[fill] (v\k\l) circle (0.02);
      }
    }

    \draw[thick, black, line join=miter] (v00) -- (v10) -- (v20) -- (v21) -- (v22) -- (v12) -- (v02) -- (v01) -- cycle;
    \draw[thick, black, line join=miter] (v01) -- (v11) -- (v21);
    \draw[thick, black, line join=miter] (v10) -- (v11) -- (v12);
    \draw[thick, black, line join=miter] (v01) -- (v10);
    \draw[thick, black, line join=miter] (v02) -- (v11) -- (v20);
    \draw[thick, black, line join=miter] (v12) -- (v21);

    \foreach \i/\j in {0/1, 1/2} {
      \foreach \k in {0, 1, 2} {
        \draw[fill] (barycentric cs:v\i\k=1,v\j\k=1) circle (0.02);
        \draw[fill] (barycentric cs:v\k\i=1,v\k\j=1) circle (0.02);
      }
    }

    \draw[fill] (barycentric cs:v10=1,v01=1) circle (0.02);
    \draw[fill] (barycentric cs:v21=1,v12=1) circle (0.02);
    \draw[fill] (barycentric cs:v20=1,v11=1) circle (0.02);
    \draw[fill] (barycentric cs:v11=1,v02=1) circle (0.02);
  \end{tikzpicture}

  \caption{Patch of cells for Vanka relaxation applied to a
    Taylor--Hood discretisation. The velocity degrees of freedom
    on the boundary of the patch are included, but the pressure degrees
    of freedom on the boundary are excluded.\label{fig:vanka-taylor-hood}}
\end{figure}
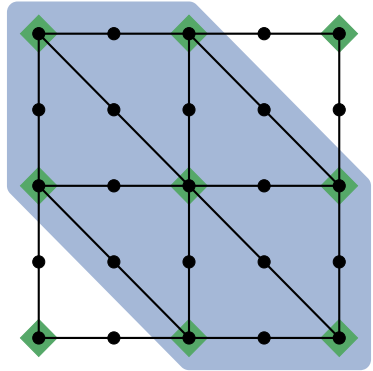

Variants of the cellwise analogue, solving simultaneously for all degrees of freedom in a
cell, have also been proposed many times in the literature \citep{fischer1997,bastian2012,bastian2019}.
It was employed by Vanka as the relaxation in a nonlinear monolithic
multigrid method for a marker-and-cell discretisation of the Navier--Stokes equations with piecewise
constant pressures \citep{vanka1986}. This idea can be generalised to other
discretisations of saddle point problems by constructing patches that gather all
degrees of freedom connected to a single degree of freedom for the Lagrange
multiplier \citep{maclachlan2011}. For example, Vanka relaxation applied to a Taylor--Hood
discretisation of the Navier--Stokes equations (with continuous piecewise
linear pressures) builds patches around each
vertex as shown in \cref{fig:vanka-taylor-hood}, including the velocity (but not pressure) degrees of freedom on the boundary of the
cells around the vertex.
This is a larger patch than the vertex-star.

In general, these relaxation methods can be viewed as additive or multiplicative
Schwarz methods induced by a space decomposition \citep{xu1992}
\begin{equation}
  \label{eq:abstract-decomposition}
V = \sum_i V_i,
\end{equation}
where $V$ is the trial space of the discretisation and each $V_i$ is constructed
by gathering the degrees of freedom associated with a given subset of
topological entities in the mesh\footnote{By topological entity we
  mean a vertex, edge, face, or cell of the mesh.}.
The abstract formulation~\eqref{eq:abstract-decomposition} also
includes classical domain decomposition methods (when augmented with
appropriate transmission conditions)
\citep{toselli2005,smith1996,chan1994c,dolean2015}. Classical domain
decomposition uses a relatively small number of large subspaces,
whereas we are interested in making the subspaces as small as
possible, while remaining effective for eliminating the relevant component of the error
in a multigrid cycle.

There exists a large amount of software for classical domain
decomposition \citep{hecht2012,jolivet2013,zampini2016}, but due to
this difference in patch size, the available software is not always well-suited for use as a relaxation
method in a multigrid context. As a consequence, despite optimal
multigrid relaxation methods being known for many hard problems, there
are few (if any) general software implementations that allow easy
access to them. For example, a Reynolds-robust multigrid scheme for
the Navier--Stokes equations was developed in \citet{benzi2006}, but
no general implementation of the full scheme was available until the work of
\citet{farrell2018b} (based on \pcpatch).

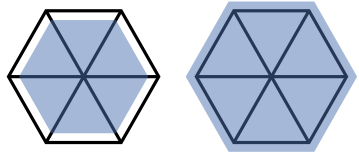
\begin{figure}[htbp]
  \centering
  \begin{tikzpicture}
    \coordinate(0-0) at (0, 0);
    \coordinate(1-0) at (1, 0);
    \coordinate(0-1) at ($(0-0) + (120:1)$);
    \coordinate(1-1) at ($(0-0) + (60:1)$);
    \coordinate(2-1) at ($(1-0) + (60:1)$);
    \coordinate(0-2) at ($(0-1) + (60:1)$);
    \coordinate(1-2) at ($(1-1) + (60:1)$);

    \draw[line join=miter, very thick] (0-0) -- (1-0) -- (2-1) -- (1-2)
    -- (0-2) -- (0-1) -- cycle;

    \draw[line join=miter, very thick] (0-0) -- (1-1) -- (1-0);
    \draw[line join=miter, very thick] (2-1) -- (1-1) -- (1-2);
    \draw[line join=miter, very thick] (0-2) -- (1-1) -- (0-1);

    \path[fill=blue, opacity=0.5] ([shift={(60:4pt)}]0-0) --
    ([shift={(120:4pt)}]1-0) --
    ([shift={(180:4pt)}]2-1) --
    ([shift={(240:4pt)}]1-2) --
    ([shift={(300:4pt)}]0-2) --
    ([shift={(0:4pt)}]0-1) --
    cycle;
    \begin{scope}[xshift=2.5cm]
      \coordinate(0-0) at (0, 0);
      \coordinate(1-0) at (1, 0);
      \coordinate(0-1) at ($(0-0) + (120:1)$);
      \coordinate(1-1) at ($(0-0) + (60:1)$);
      \coordinate(2-1) at ($(1-0) + (60:1)$);
      \coordinate(0-2) at ($(0-1) + (60:1)$);
      \coordinate(1-2) at ($(1-1) + (60:1)$);

      \draw[line join=miter, very thick] (0-0) -- (1-0) -- (2-1) -- (1-2)
      -- (0-2) -- (0-1) -- cycle;

      \draw[line join=miter, very thick] (0-0) -- (1-1) -- (1-0);
      \draw[line join=miter, very thick] (2-1) -- (1-1) -- (1-2);
      \draw[line join=miter, very thick] (0-2) -- (1-1) -- (0-1);

      \path[fill=blue, opacity=0.5] ([shift={(60:-4pt)}]0-0) --
      ([shift={(120:-4pt)}]1-0) --
      ([shift={(180:-4pt)}]2-1) --
      ([shift={(240:-4pt)}]1-2) --
      ([shift={(300:-4pt)}]0-2) --
      ([shift={(0:-4pt)}]0-1) --
      cycle;
    \end{scope}
  \end{tikzpicture}
  \caption{The star of a vertex (left), and its closure (right).\label{fig:starclosure}}
\end{figure}
\pcpatch constructs the subspaces as follows. Sets of topological
entities are gathered according to a user-specified rule. For example,
the vertex-star iteration loops over all vertices of the mesh and
applies the \emph{star operation} to gather a set of entities for each vertex.
The star operation is a standard concept in algebraic topology: given
a simplicial complex $K$, the star of a simplex $p$ is the union of
the interiors of all simplices that contain $p$ \citep[\S
2]{munkres1984}. Concretely, the star of a vertex is the interiors of all edges,
faces, and cells incidental to the vertex, along with the vertex
itself, as shown on the left of \cref{fig:starclosure}. To define each subspace $V_i$ from these sets of entities we
gather all degrees of freedom supported on the entities, and set
$V_i$ to be the span of the associated basis functions.

On each $V_i$ a small linear or non-linear problem must be solved. To
do this, \pcpatch determines the cells involved in the assembly of
this problem by an adjacency relation appropriate to the
discretisation. For example, standard finite element assembly
involving only cell integrals requires the \emph{closure} of the
entities in $V_i$. The closure of a set of simplices is the smallest
simplicial subcomplex that contains each simplex in the set. Concretely, in the
vertex-star example, this means adding all edges, vertices, and faces
of cells in the set, i.e.~including the boundary, as shown on the right
of \cref{fig:starclosure}. We refer to this final set of
entities as a \emph{patch}. \pcpatch then invokes a user-provided
callback function to assemble the matrix (and residual if nonlinear) of
the problem on each cell of the patch. The local problem is then
assembled using an appropriate local numbering, and solved (often with
a direct factorisation, but not necessarily). The solution to the
local problem results in a local update to the global solution that
\pcpatch applies either additively or multiplicatively, as selected
by the user.

The advantage of organising the software in this way is its great
flexibility. For example, to implement Vanka relaxation for
the marker-and-cell discretisation, the only change required is to
specify that patches are created by iterating over cells using the
closure operation to gather the entities in each patch; everything
else follows automatically. To implement Vanka relaxation for
Taylor--Hood, the only change required is to specify that patches are
created by iterating over vertices using the closure of the star to
gather entities in each patch. The callback structure separates the
specification of a space decomposition from the implementation of
the discretisation. Consequently, \pcpatch requires no new code to be
written when attacking a problem in a new way.

The remainder of the manuscript is organised as follows.
The mathematical theory of space decompositions and subspace
corrections is reviewed in \cref{sec:math-backgr},
with a particular emphasis on achieving parameter-robust
convergence.
The resulting computational abstractions developed in \pcpatch
are described in \cref{sec:comp-abstr}. The software
is applied to several challenging problems in \cref{sec:examples},
demonstrating its flexibility and utility. We conclude in \cref{sec:conclusion}.
\section{Mathematical background}\label{sec:math-backgr}

For ease of exposition, we restrict ourselves in this section to
linear variational problems, although \pcpatch also applies to
nonlinear problems. Consider the problem: find $u \in V$ such that
\begin{equation}
  \label{eq:1}
  a(u, v) = (f, v) \text{  for all $v \in V$},
\end{equation}
where $V$ is a finite-dimensional Hilbert space, $a : V \times V \to
\mathbb{R}$ is a bilinear form, and $f$ is right hand side data in an
appropriate function space. Many algorithms for
solving~\eqref{eq:1} are induced by a space decomposition of $V$ into
subspaces $V_i$ with
\begin{equation}
  \label{eq:2}
  V = \sum_{i=1}^{J} V_i.
\end{equation}
This means that for any $v \in V$ there exists $\{v_i\}_{i=1}^J$ with
$v = \sum_{i=1}^J v_i$ and $v_i \in V_i$. This decomposition is
usually not unique.

There are two main solver algorithms associated with this space
decomposition. The additive variant, referred to as parallel subspace
correction in \citet{xu1992}, solves for updates to the solution in
each subspace independently, and is shown in \cref{alg:psc}. Given an initial guess $u^k
\in V$, the associated error equation is: find $e \in V$ such that
\begin{equation}
  \label{eq:4}
  a(e, v) = (f, v) - a(u^k, v) \text{ for all } v \in V.
\end{equation}
If we could compute $e$, then $u^k + e$ would be the exact solution.
The idea of parallel subspace correction is to compute an
approximation $\delta u_i$ to $e$ in each subspace $V_i$ independently by solving
the Galerkin projection of~\eqref{eq:4} onto $V_i$\footnote{The
  analysis of \citet{xu1992} does not require that these solves are
  exact. However, in this paper we only consider exact subspace
  solves, both to simplify the exposition and because this is typically
  what is done in the context of multigrid relaxation.}. These updates
$\delta u_i$ are then combined, possibly after application of a
weighting operator (often a partition of unity) $w_i$ on each
subspace.
\begin{algorithm}[htbp]
  \SetKwInOut{Input}{input}\SetKwInOut{Output}{output}
  \Input{\,Initial guess $u^k \in V$}
  \Input{\,Weighting operators $w_i : V_i \to V_i$}
  \Output{\,Updated guess $u^{k+1} \in V$}
  \BlankLine
  \For{$i = 1$ \KwTo $J$}{
    Find $\delta u_i \in V_i$ such that
    \begin{equation*}
      a(\delta u_i, v_i) = (f, v_i) - a(u^k, v_i) \text{ for all } v_i \in V_i.
    \end{equation*}
  }
  $u^{k+1} \gets u^k + \sum_{i=1}^{J} w_i(\delta u_i)$
  \caption{Parallel subspace correction\label{alg:psc}}
\end{algorithm}

The multiplicative algorithm, referred to by \citet{xu1992} as
sequential subspace correction, solves for updates to the solution in
each subspace sequentially, and is shown in \cref{alg:ssc}. The
difference to parallel subspace correction is that the updates from
each subspace solve are immediately applied to the current guess,
which modifies the right hand side of the error equation for the next
solve. The multiplicative variant typically exhibits better
convergence, but the subsolves cannot be parallelised and the residual
must be updated at each step.
\begin{algorithm}[htbp]
  \SetKwInOut{Input}{input}\SetKwInOut{Output}{output}
  \Input{\,Initial guess $u^k \in V$}
  \Output{\,Updated guess $u^{k+1} \in V$}
  \BlankLine
  \For{$i = 1$ \KwTo $J$}{
    Find $\delta u_i \in V_i$ such that
    \begin{equation*}
      a(\delta u_i, v_i) = (f, v_i) - a(u^{k + (i-1)/J}, v_i) \text{ for all } v_i \in V_i.
    \end{equation*}
    $u^{k+i/J} \gets u^{k + (i-1)/J} + \delta u_i$
  }
  \caption{Sequential subspace correction\label{alg:ssc}}
\end{algorithm}

Given basis functions $\{\phi_1, \dots, \phi_N\}$ for $V$, the
classical Jacobi and Gau\ss--Seidel iterations are induced by the
space decomposition
\begin{equation}
  \label{eq:5}
  V = \sum_{i=1}^{N} \operatorname{span} \{\phi_i\}
\end{equation}
and by the additive and multiplicative algorithms respectively.

A domain decomposition method is induced by the space decomposition
\begin{equation}
  \label{eq:6}
  V = V_0 + \sum_{i=1}^{J} V_i,
\end{equation}
where $V_0$ is a coarse space and $V_i$ consists of functions
supported in some subdomain $\Omega_i$ of the domain $\Omega \subset
\mathbb{R}^d$. Typically, the updates are combined with the additive
algorithm\footnote{In the domain decomposition context, the problem
  solved on each subspace is often modified, for example by imposing
  boundary conditions other than homogeneous Dirichlet. See
  \citet{gander2006} for details.}.

A multigrid V-cycle is induced by the multiplicative algorithm applied to
the space decomposition
\begin{equation}
  \label{eq:7}
  V = \sum_{l=L}^2 V_l + V_1 + \sum_{l=2}^{L} V_l,
\end{equation}
where $V_1 \subset V_2 \subset \dots \subset V_L = V$. Typically, each
$V_l$ is constructed on the refinement of the mesh used for $V_{l-1}$. The updates are
performed with
approximate solvers for $l > 1$. In fact, the approximate solver on each level $l > 1$
is often itself a subspace correction method.

Of course, the effectiveness of these solvers depends on the space decomposition chosen.
We briefly recall some of the standard theory for subspace correction methods that describes
what makes for an effective space decomposition.
We define the operator $A:V\to V^*$ associated with the bilinear form via
\begin{equation}
    \langle A u, v\rangle = a(u, v) \quad \text{ for all } u, v \in V,
\end{equation}
where $V^*$ is the dual space of $V$.
For each subspace we denote the inclusion $I_i : V_i \to V$ and its adjoint $I_i^* : V^* \to V_i^*$, and we define the restriction of $A$ to $V_i$ by
\begin{equation}
    \langle A_i u_i, v_i \rangle = \langle A I_i u_i, I_i v_i \rangle \quad \text{ for all } u_i, v_i \in V_i,
\end{equation}
i.e.~$A_i = I_i^* A I_i$.
We assume that the patch solves are performed exactly.
The parallel subspace correction preconditioner associated with the decomposition $\{V_i\}$ can then be expressed as
\begin{equation}
    D^{-1} = \sum_i I_i A_i^{-1} I_i^*.
\end{equation}
Denote $T=D^{-1} A=\sum_i T_i$, where $T_i = I_i A_i^{-1} I_i^* A$.
Assuming that $a$ is symmetric, bounded, and coercive and that $D^{-1}$ will be
used as a preconditioner in the conjugate gradient algorithm,
then the goal is to estimate the condition number $\kappa(T)$ bounding
the convergence of the Krylov method.
\begin{theorem}
   Assume that there exist constants $c_0$ and $c_1$ such that
   \begin{equation}\label{eqn:interaction}
       (Tu, Tu)_A \le c_0 (Tu, u)_A
   \end{equation}
   and
   \begin{equation}\label{eqn:splitting}
       \inf_{\substack{u_i\in V_i\\\sum_i u_i = u}}\sum_i\|u_i\|_{A_i}^2 \le c_1 \|u\|_A^2
   \end{equation}
   for all $u \in V$.
   Then
   \begin{equation}
       \lambda_\mathrm{min}(T) \ge c_1^{-1} \quad \text{ and } \quad \lambda_\mathrm{max}(T) \le c_0.
     \end{equation}
\end{theorem}
A proof can be found in~\citet[Theorem 4.1]{xu1992}.
We briefly comment on the two conditions.
The first condition measures the interaction of the subspaces $V_i$.
A useful tool is the interaction matrix $\Theta$ with entries $\Theta_{ij}$ defined by the smallest constants satisfying
\begin{equation}
    |( T_i u, T_j v )_A| \le \Theta_{ij} \big[(T_i u, u )_A\big]^{1/2} \big[( T_j v, v )_A\big]^{1/2} \quad \forall u, v\in V.
\end{equation}
One can show that
\begin{equation}
    c_0 \le \rho(\Theta)
\end{equation}
where $\rho(\Theta)$ is the spectral radius of $\Theta$ (\citet[Lemma 4.6]{xu1992}).
It follows that
\begin{equation}
    \rho(\Theta) \le \|\Theta\|_1 = \max_j \sum_i |\Theta_{ij}| \le N_O,
\end{equation}
where $N_O$ is the maximal number of overlapping subspaces of any one subspace.
In particular, this shows that the interaction can be estimated by
purely topological arguments: in the case of the vertex-star on a
regular mesh as shown in \cref{fig:hcurl-patch}, the interaction can be
bounded by $c_0 \le N_O = 7$. In general, $N_O$ is bounded on shape
regular meshes.

The constant $c_1$ estimates the stability of the space decomposition and the eigenvalue bound follows from the classic statement
\begin{equation}
    \|u\|_D^2 = \inf_{\substack{u_i\in V_i\\\sum_i u_i = u}} \sum_i \|u_i\|_{A_i}^2,
\end{equation}
found in \citet[eqn.~(4.11)]{xu2001}.
We illustrate the calculation of $c_1$ in the next section.

\subsection{Parameter-robust subspace correction}\label{sec:multigrid-relaxation-sc}

The problem~\eqref{eq:1} to be solved often depends on a parameter
$\alpha \in \mathbb{R}$. For example, the linearised Navier--Stokes operator
depends on the Reynolds number, while the equations of linear
elasticity depend on the Poisson ratio. It is desirable to build
multigrid methods with parameter-robust convergence, i.e.~the
number of iterations required for convergence does not vary
substantially as $\alpha$ is varied. A requirement for
achieving this is that the relaxation method is also parameter-robust.
In some important cases, the construction of parameter-robust relaxation methods
can be achieved by an appropriate choice of space decomposition. In
particular, for nearly singular symmetric problems and when $\alpha \ge 0$, the appropriate
choice is guided by results of \citet{schoberl1999,schoberl1999b} (for
the additive case) and \citet{lee2007} (for the multiplicative case).

Before we state the results, we consider a simple example.
Let $V$ be an $H^1_0$-conforming finite element space, and let
\begin{equation}
    a(u,v) = (\nabla u, \nabla v) + \alpha (\nabla\cdot u, \nabla \cdot v),
\end{equation}
for $\alpha \ge 0$.
In the simple case of Jacobi smoothing, then the decomposition $u=\sum u_i$ is unique and using a standard inverse estimate we obtain
\begin{equation}
    \begin{aligned}
        \|u\|_{D}^2 &= \sum_i \|u_i\|_{A}^2 \preceq (1+\alpha) \sum_i \|u_i\|_1^2 \preceq \frac{1+\alpha}{h^2} \sum_i \|u_i\|_0^2 \\
                            &\preceq (1+\alpha) h^{-2} \|u\|_0^2 \preceq (1+\alpha) h^{-2} \|u\|_{A}^2,
    \end{aligned}
\end{equation}
and hence $c_1 \sim (1+\alpha) h^{-2}$.

We make two observations.
First, the bound grows as the mesh is refined.
This is well known and is the reason why Jacobi relaxation by itself is not effective but must be used in a multigrid hierarchy.
Second, the bound increases with $\alpha$: Jacobi relaxation is not parameter-robust.
To avoid this blow-up, the space decomposition needs to \emph{decompose the nullspace of the singular operator} that is scaled with $\alpha$.

This argument was made rigorous by Sch\"oberl \citep[Theorem 4.1]{schoberl1999b}
for the case of the parallel subspace correction method, and a similar
result also holds for the sequential subspace correction method
\citep[Theorem 4.2]{lee2007}.

\begin{theorem}[Parameter robust parallel subspace correction]\label{thm:kernel-decomp}
    Consider a problem of the form: for $\alpha \ge 0$, find $u \in V$ such that
    \begin{equation}
        a(u,v) =  (f, v) \quad \text{ for all } v \in V,
    \end{equation}
    where
    \begin{equation}
    a(u, v) = a_0(u, v) + \alpha b(\Lambda u, \Lambda v)
    \end{equation}
    and $\Lambda: V\to Q$ for some space $Q$.
    Assume that the mapping $\Lambda$ is a continuous linear map and that $b$ is symmetric, bounded, and coercive on $Q$ and that $a_0$ is symmetric, bounded, and coercive on $V$.
    Denote the kernel by
    \begin{equation}
    \mathcal{N} = \{ u \in V \colon b(\Lambda u, q) = 0 \, \forall\, q \in Q\}.
    \end{equation}
    A space decomposition
    \begin{equation}
        V = \sum_i V_i
    \end{equation}
    defines a subspace correction method that is robust with respect to
    $\alpha$ if in addition to~\eqref{eqn:interaction} and~\eqref{eqn:splitting} the decomposition satisfies the following properties:
    \begin{enumerate}
        \item the pair $V\times Q$ is inf-sup stable for the mixed problem induced by
            \begin{equation}
                B((u, p), (v, q)) = a_0(u, v) - b(\Lambda u, q) - b(\Lambda v, p),
            \end{equation}
          \item the splitting is stable for the $V$-norm:
            \begin{equation}
              \label{eqn:splitting-a0}
              \inf_{\substack{u_i\in V_i \\\sum_i u_i = u}}\sum_i\|u_i\|_{V}^2 \le c_1 \|u\|_{V}^2 \quad \text{ for all } u\in V
            \end{equation}
    for some $c_1$,
        \item the splitting is stable in the operator norm on the kernel $\mathcal N$:
   \begin{equation}\label{eqn:splitting-nullspace}
       \inf_{\substack{u_i\in V_i\cap \mathcal{N} \\\sum_i u_i = u}}\sum_i\|u_i\|_{A_i}^2 \le c_2 \|u\|_A^2 \quad \text{ for all } u\in V\cap\mathcal{N}
   \end{equation}
    for some $\alpha$-independent $c_2$.
    \end{enumerate}
\end{theorem}

We highlight that in particular the condition~\eqref{eqn:splitting-nullspace} requires that
\begin{equation}
    \label{eqn:lee2007}
    \mathcal{N} = \sum_i (V_i \cap \mathcal{N}).
\end{equation}
In other words, a crucial property for such nearly singular systems
is that any kernel function can be written as a sum of kernel
functions in the subspaces.

We give an example of the construction of subspaces $\{V_i\}$ so that~\eqref{eqn:lee2007} is satisfied for the $H(\curl)$ Riesz map in three dimensions~\eqref{eq:hcurl}.
We consider the de Rham complex and the Whitney subcomplex
\begin{equation}
  \label{eq:9}
\begin{tikzcd}
  H^1 \arrow[r, "\grad"] \arrow[d]  & H(\curl) \arrow[r, "\curl"]
  \arrow[d] & H(\div) \arrow[r, "\div"]  \arrow[d] & L^2  \arrow[d] \\
  \Sigma \arrow[r, "\grad"] & V \arrow[r, "\curl"] & W \arrow[r, "\div"] & Q
\end{tikzcd}.
\end{equation}
Here $\Sigma$ is the familiar space of continuous Lagrange finite elements of order $k$, $V$ is the N\'ed\'elec finite element of the first kind of order $k$, $W$ is the Raviart--Thomas element of order $k$, and $Q$ is the space of discontinuous Lagrange finite elements of order $k-1$.
This sequence is exact on simply-connected domains, and hence for every $u\in H(\curl)$ with $\nabla\times u = 0$, we can find a $\Phi\in H^1$ such that $\nabla\Phi= u$.
This property carries over to the discrete complex:
for $u\in V$ with $\nabla\times u =0$ there exists a $\Phi\in \Sigma$ with $\nabla\Phi = u$.
Given a basis $\{\Phi_j\}$ for $\Sigma$, we can write $\Phi = \sum_{j}\alpha_j\Phi_j$ and defining
\begin{equation}
    u_j = \alpha_i\nabla \Phi_j,
\end{equation}
we observe that
\begin{equation}
    \sum_j u_j = u \quad \text{ and } \quad \nabla\times u_j = 0 \text{ for all } j.
\end{equation}
We now need to find subspaces $\{V_i\}$ such that for every $u_j$ there exists a $V_i$ with $u_j\in V_i$.
To do this, we examine the support of the basis functions.
Since the standard finite element basis for $\Sigma$ has basis functions supported in the stars of vertices, the same holds for each $u_j$, as it is simply the gradient of such a function.
Consequently, defining $V_i$ to be the functions supported in the star of vertex $i$ gives the desired kernel capturing property~\eqref{eqn:lee2007}.

Obtaining the bound in~\eqref{eqn:splitting-nullspace} is harder and requires bounds on the potential $\Phi$ in terms of $u$.
We note that such bounds are usually known for the infinite-dimensional sequence and also hold for the discrete sequence when commuting projections exist.
For more detail on multigrid relaxation methods derived from exterior calculus we refer to~\citet{arnold2006}.

\subsection{Nonlinear relaxation}
The idea of subspace correction also applies to the nonlinear case. Consider the problem:
find $u \in V$ such that
\begin{equation} \label{eqn:nonlin}
F(u; v) = 0 \text{ for all } v \in V,
\end{equation}
where $F: V \times V \to \mathbb{R}$ is the residual. Given an initial
guess $u^k$, the associated error equation is to find $e \in V$ such
that $F(u^k + e; v) = 0$ for all $v \in V$.  The associated additive
algorithm that computes approximations to $e$ in each subspace is given
in \cref{alg:pscnonlin}.
\begin{algorithm}[htbp]
  \SetKwInOut{Input}{input}\SetKwInOut{Output}{output}
  \Input{\,Initial guess $u^k \in V$}
  \Input{\,Weighting operators $w_i : V_i \to V_i$}
  \Output{\,Updated guess $u^{k+1} \in V$}
  \BlankLine
  \For{$i = 1$ \KwTo $J$}{
    Find $\delta u_i \in V_i$ such that
    \begin{equation*}
        F(u^k + \delta u_i; v_i) = 0 \text{ for all } v_i \in V_i.
    \end{equation*}
  }
  $u^{k+1} \gets u^k + \sum_{i=1}^{J} w_i(\delta u_i)$
  \caption{Parallel subspace correction for the nonlinear problem~\eqref{eqn:nonlin}.\label{alg:pscnonlin}}
\end{algorithm}
Initial theoretical results on this algorithm were established by
\citet{dryja1997}.
This algorithm is often used as a nonlinear preconditioner for an outer Newton method \citep{cai2002,brune2015}.

We note that in the context of finite elements the residual $F$ usually has a local nature and when testing with functions in a patch spanning the space $V_i$, only the part of $u^k$ that overlaps with $V_i$ is relevant.
To formalise this property we assume that there exist subspaces $\overline{V_i}\supset V_i$, injection operators $\iota_i:V\to \overline{V_i}$, and local residuals $F_i : \overline{V_i}\to V_i^*$ such that 
\begin{equation}
    F(u^k +\delta u_i; v_i) = F_i(\iota (u^k) + \delta u_i; v_i) \quad \text{ for all } v_i \in V_i,\ \delta u_i \in V_i.
\end{equation}
This means that the local Newton solvers do not need to know the full state, but only the relevant part of the state given by $\iota_i(u^k)$; this is important for parallelisation. As a consequence, in the implementation we solve for $\delta u_i\in V_i$ such that
\begin{equation}
    F_i(\iota_i(u^k) + \delta u_i; v_i) = 0 \text{ for all } v_i \in V_i.
\end{equation}


\section{Computational abstractions}\label{sec:comp-abstr}

Even though mathematical formulations of parameter robust multigrid
relaxation schemes have been known for several decades,
implementations of these schemes have been missing from general
purpose linear algebra and discretisation packages such as PETSc
\citep{petsc-user-ref}, Trilinos \citep{heroux2005}, deal.II
\citep{bangerth2007}, DUNE \citep{blatt2016}, Firedrake
\citep{Rathgeber:2016}, and FEniCS \citep{Logg2010,Logg2012}, due to
the difficulty of defining the space decomposition and resulting local
problems in a general and composable way. A popular algebraic approach
for implementing subspace correction methods, such as provided by
PETSc in \texttt{\href{https://www.mcs.anl.gov/petsc/petsc-current/docs/manualpages/PC/PCASM.html}{\underline{PCASM}}}, is to obtain index sets that define the space
decomposition and construct the local operators algebraically by
extracting submatrices from the global assembled matrix. NGSolve
\citep{Schoeberl2014} also
supports this algebraic approach, augmented with the ability to obtain
the index sets through
topological queries \citep{ngsolveweb2017}.  Although
successful, using such an interface to construct multigrid relaxation
schemes has several disadvantages. It does not allow for matrix-free
implementations (which can offer significant efficiency advantages at
high order) or efficient nonlinear relaxation (since each local
nonlinear update would require the assembly of the global Jacobian).
In light of these issues, we take a different approach: separating the
topological decomposition and subproblem assembly into different
stages, and using a callback interface that we now describe.

\pcpatch separates the assembly of subproblems on patches into three
stages. First we decompose the mesh into patches of entities whose
degrees of freedom will be updated simultaneously in each subproblem.
For this we use PETSc's \texttt{\href{https://www.mcs.anl.gov/petsc/petsc-current/docs/manualpages/DMPLEX/DMPLEX.html}{\underline{DMPLEX}}} mesh abstraction
\citep{Knepley:2005,Knepley:2009,lange2015}, which gives us a simple
dimension-independent topological query language. For example, it
offers the star and closure operations that often arise when describing
patches. Second, we gather the degrees
of freedom in the function space associated to the entities in each
patch and build local numberings for restrictions of the function
space to the patch. Third, these numberings are supplied to callbacks
to a discretisation engine\footnote{Currently Firedrake and PETSc's
  \texttt{PetscDS} are supported; interfacing to other
  discretisation engines is possible.} to conduct the assembly of the
subproblem (Jacobian and possibly residual) on the patch. The
callbacks receive the current local state, a list of entities in the
global mesh to assemble over, and the degree of freedom mappings for
the local function space. The C signatures of these callbacks are
shown in \cref{lst:assembly-callbacks}. By separating the assembly of
patches in this way, the abstraction is robust to the type of mesh
and element employed, the PDE being solved, and the global numbering
of degrees of freedom.
\begin{listing}[htbp]
\centering
\newcommand{\jaclink}{https://www.mcs.anl.gov/petsc/petsc-current/docs/manualpages/PC/PCPatchSetComputeOperator.html}
\newcommand{\functionlink}{https://www.mcs.anl.gov/petsc/petsc-current/docs/manualpages/PC/PCPatchSetComputeFunction.html}
\begin{minted}[escapeinside=@@]{c}
int @\href{\jaclink}{\underline{\texttt{ComputeJacobian}}}@(PC pc, PetscInt point, Vec state, Mat J, IS cells,
  PetscInt ndof, PetscInt *dofNumbering, PetscInt *dofNumberingWithBoundary);

int @\href{\functionlink}{\underline{\texttt{ComputeResidual}}}@(PC pc, PetscInt point, Vec state, Vec F, IS cells,
  PetscInt ndof, PetscInt *dofNumbering, PetscInt *dofNumberingWithBoundary);
\end{minted}
\caption{Function signatures of the Jacobian and residual callbacks
for assembly of subproblems on each patch.\label{lst:assembly-callbacks}}
\let\jaclink\relax
\let\functionlink\relax
\end{listing}

The data structures representing the local function spaces are deliberately
kept lightweight. In particular, no mesh object representing the subdomain
is built. For example, information about which facets are on the
boundary of the patch is not stored.
This is in contrast to the domain decomposition approach used
in the high-level FreeFem++ library \citep{hecht2012}, which does build
fully-fledged meshes on each subdomain. Building submeshes facilitates code re-use (no
callback interface is required) and is efficient when $\mathcal{O}(1)$
subdomains per MPI process are employed, but is computationally expensive in time
and memory in the multigrid context where many small subdomains are expected.
The strategy we use is to mark the patch entities within the larger
mesh. The discretisation engine can use this information to query the
global mesh if necessary.
This allows rapid assembly without the overhead of object creation while
still allowing us access to topological queries so that we can, for example,
construct the patch boundary on the fly.

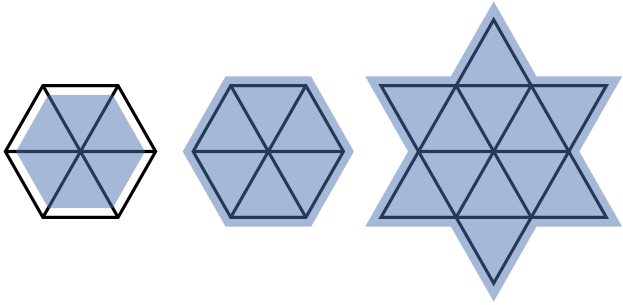
\begin{figure}[htbp]
  \centering
  \begin{tikzpicture}
    \coordinate(0-0) at (0, 0);
    \coordinate(1-0) at (1, 0);
    \coordinate(0-1) at ($(0-0) + (120:1)$);
    \coordinate(1-1) at ($(0-0) + (60:1)$);
    \coordinate(2-1) at ($(1-0) + (60:1)$);
    \coordinate(0-2) at ($(0-1) + (60:1)$);
    \coordinate(1-2) at ($(1-1) + (60:1)$);

    \draw[line join=miter, very thick] (0-0) -- (1-0) -- (2-1) -- (1-2)
    -- (0-2) -- (0-1) -- cycle;

    \draw[line join=miter, very thick] (0-0) -- (1-1) -- (1-0);
    \draw[line join=miter, very thick] (2-1) -- (1-1) -- (1-2);
    \draw[line join=miter, very thick] (0-2) -- (1-1) -- (0-1);

    \path[fill=blue, opacity=0.5] ([shift={(60:4pt)}]0-0) --
    ([shift={(120:4pt)}]1-0) --
    ([shift={(180:4pt)}]2-1) --
    ([shift={(240:4pt)}]1-2) --
    ([shift={(300:4pt)}]0-2) --
    ([shift={(0:4pt)}]0-1) --
    cycle;
    \begin{scope}[xshift=2.5cm]
      \coordinate(0-0) at (0, 0);
      \coordinate(1-0) at (1, 0);
      \coordinate(0-1) at ($(0-0) + (120:1)$);
      \coordinate(1-1) at ($(0-0) + (60:1)$);
      \coordinate(2-1) at ($(1-0) + (60:1)$);
      \coordinate(0-2) at ($(0-1) + (60:1)$);
      \coordinate(1-2) at ($(1-1) + (60:1)$);

      \draw[line join=miter, very thick] (0-0) -- (1-0) -- (2-1) -- (1-2)
      -- (0-2) -- (0-1) -- cycle;

      \draw[line join=miter, very thick] (0-0) -- (1-1) -- (1-0);
      \draw[line join=miter, very thick] (2-1) -- (1-1) -- (1-2);
      \draw[line join=miter, very thick] (0-2) -- (1-1) -- (0-1);

      \path[fill=blue, opacity=0.5] ([shift={(60:-4pt)}]0-0) --
      ([shift={(120:-4pt)}]1-0) --
      ([shift={(180:-4pt)}]2-1) --
      ([shift={(240:-4pt)}]1-2) --
      ([shift={(300:-4pt)}]0-2) --
      ([shift={(0:-4pt)}]0-1) --
      cycle;
    \end{scope}
    \begin{scope}[xshift=5.5cm]
      \coordinate(0-0) at (0, 0);
      \coordinate(1-0) at (1, 0);
      \coordinate(0-1) at ($(0-0) + (120:1)$);
      \coordinate(1-1) at ($(0-0) + (60:1)$);
      \coordinate(2-1) at ($(1-0) + (60:1)$);
      \coordinate(0-2) at ($(0-1) + (60:1)$);
      \coordinate(1-2) at ($(1-1) + (60:1)$);

      \coordinate(1--1) at ($(0-0) + (-60:1)$);
      \coordinate(-1-0) at ($(0-0) + (180:1)$);
      \coordinate(2-0) at ($(1-0) + (0:1)$);
      \coordinate(-1-2) at ($(0-2) + (180:1)$);
      \coordinate(2-2) at ($(1-2) + (0:1)$);
      \coordinate(1-3) at ($(0-2) + (60:1)$);

      \draw[line join=miter, very thick] (0-0) -- (1-0) -- (2-1) -- (1-2)
      -- (0-2) -- (0-1) -- cycle;

      \draw[line join=miter, very thick] (0-0) -- (1-1) -- (1-0);
      \draw[line join=miter, very thick] (2-1) -- (1-1) -- (1-2);
      \draw[line join=miter, very thick] (0-2) -- (1-1) -- (0-1);

      \draw[line join=miter, very thick] (0-0) -- (1--1) -- (1-0);
      \draw[line join=miter, very thick] (0-0) -- (-1-0) -- (0-1);
      \draw[line join=miter, very thick] (0-1) -- (-1-2) -- (0-2);
      \draw[line join=miter, very thick] (0-2) -- (1-3) -- (1-2);
      \draw[line join=miter, very thick] (1-2) -- (2-2) -- (2-1);
      \draw[line join=miter, very thick] (2-1) -- (2-0) -- (1-0);

      \path[fill=blue, opacity=0.5] ([shift={(-120:4pt)}]0-0) --
      ([shift={(-90:6.928203pt)}]1--1) -- 
      ([shift={(-60:4pt)}]1-0) --
      ([shift={(-30:6.928203pt)}]2-0) --
      ([shift={(0:4pt)}]2-1) --
      ([shift={(30:6.928203pt)}]2-2) --
      ([shift={(60:4pt)}]1-2) --
      ([shift={(90:6.928203pt)}]1-3) --
      ([shift={(120:4pt)}]0-2) --
      ([shift={(150:6.928203pt)}]-1-2) --
      ([shift={(180:4pt)}]0-1) --
      ([shift={(-150:6.928203pt)}]-1-0) --
      cycle;
    \end{scope}
  \end{tikzpicture}
  \caption{Vertex-star patch (left), its completion when the problem
    requires only cell integrals (centre), and completion when facet
    integrals are also present (right).\label{fig:patch-completion}}
\end{figure}

Once each patch of entities defining the degrees of freedom to be
updated in a patch solve is constructed, we additionally gather all
entities in the stencil of the operator to be assembled on the patch.
We term this step \emph{patch completion}. For example, if the
operator involves integration over cells of the mesh, we extend the
patch to include all entities which lie in the closure of the cells we
integrate over. If the operator contains facet terms, the patch
completion is more involved, gathering all entities in the closure of
the support of facets in the patch. This process is illustrated in \cref{fig:patch-completion}.

For the solution of the subproblems on each patch, we use a PETSc
\texttt{\href{https://www.mcs.anl.gov/petsc/petsc-current/docs/manualpages/KSP/KSP.html}{\underline{KSP}}} (for linear problems) and \texttt{\href{https://www.mcs.anl.gov/petsc/petsc-current/docs/manualpages/SNES/SNES.html}{\underline{SNES}}} (for nonlinear
problems). This enables full flexibility in configuring the patch
solves, for example the use of inexact inverses using iterative methods. For the
common case where the patch is small and an exact inverse is desired,
we implement a small amount of special-case code to explicitly form
the inverse on each patch in setup and then solve the subproblems by
matrix-vector multiplication. This offers substantial speedups for
many problems.

\begin{figure}[htbp]
  \centering
  \begin{tikzpicture}[scale=3,
    on each segment/.style={
      decorate,
      decoration={
        show path construction,
        moveto code={},
        lineto code={
          \path [#1]
          (\tikzinputsegmentfirst) -- (\tikzinputsegmentlast);
        }}},
    nedof/.style={postaction={decorate, decoration={markings,
          mark=at position 0.5 with {
            \node[transform shape,shape=diamond,fill,draw, inner
            sep=0pt, minimum height=4pt, minimum width=8pt] {};
          }}}}]
    \foreach \i/\k in {0/0, 0.5/1, 1/2} {
      \foreach \j/\l in {0/0, 0.5/1, 1/2} {
        \coordinate (v\k\l) at (\i, \j);
      }
    }

    \draw[thick, black, line join=miter] (v00)
    -- (v10) -- (v20) -- (v21) -- (v22) -- (v12) -- (v02) -- (v01)
    -- (v00);
    \draw[nedof, thick, black, line join=cap] (v01) -- (v11);
    \draw[nedof, thick, black, line join=cap] (v11) -- (v21);
    \draw[nedof, thick, black, line join=cap] (v10) -- (v11);
    \draw[nedof, thick, black, line join=cap] (v11) -- (v12);
    \draw[thick, black, line join=cap] (v01) -- (v10);
    \draw[nedof, thick, black, line join=cap] (v02) -- (v11);
    \draw[nedof, thick, black, line join=cap] (v11) -- (v20);
    \draw[thick, black, line join=cap] (v12) -- (v21);

    \coordinate (i10) at ($(v10) + (90:0.1)$);
    \coordinate (i20) at ($(v20) + (135:0.141)$);
    \coordinate (i21) at ($(v21) + (180:0.1)$);
    \coordinate (i12) at ($(v12) + (270:0.1)$);
    \coordinate (i02) at ($(v02) + (315:0.141)$);
    \coordinate (i01) at ($(v01) + (0:0.1)$);
    \path[fill=blue, opacity=0.5]
    \convexpath{i01,i02,i12,i21,i20,i10}{1pt};

    \begin{scope}[xshift=1.5cm]
    \foreach \i/\k in {0/0, 0.5/1, 1/2} {
      \foreach \j/\l in {0/0, 0.5/1, 1/2} {
        \coordinate (v\k\l) at (\i, \j);
      }
    }
    \draw[thick, black, line join=miter] (v00) -- (v10);
    \draw[nedof, thick, black, line join=miter] (v10) -- (v20);
    \draw[nedof, thick, black, line join=miter] (v20) -- (v21);
    \draw[thick, black, line join=miter] (v21) -- (v22);
    \draw[thick, black, line join=miter] (v22) -- (v12);
    \draw[thick, black, line join=miter] (v12) -- (v02);
    \draw[thick, black, line join=miter] (v02) -- (v01);
    \draw[thick, black, line join=miter] (v01) -- (v00);

    \draw[nedof, thick, black, line join=cap] (v01) -- (v11);
    \draw[nedof, thick, black, line join=cap] (v11) -- (v21);
    \draw[nedof, thick, black, line join=cap] (v10) -- (v11);
    \draw[nedof, thick, black, line join=cap] (v11) -- (v12);
    \draw[nedof, thick, black, line join=cap] (v01) -- (v10);
    \draw[nedof, thick, black, line join=cap] (v02) -- (v11);
    \draw[nedof, thick, black, line join=cap] (v11) -- (v20);
    \draw[nedof, black, line join=cap] (v12) -- (v21);

    \coordinate (i10) at ($(v10) + (90:0.1)$);
    \coordinate (i20) at ($(v20) + (135:0.141)$);
    \coordinate (i21) at ($(v21) + (180:0.1)$);
    \coordinate (i12) at ($(v12) + (270:0.1)$);
    \coordinate (i02) at ($(v02) + (315:0.141)$);
    \coordinate (i01) at ($(v01) + (0:0.1)$);
    \path[fill=blue, opacity=0.5]
    \convexpath{i01,i02,i12,i21,i20,i10}{1pt};
    \end{scope}

    \begin{scope}[xshift=3cm]
    \foreach \i/\k in {0/0, 0.5/1, 1/2} {
      \foreach \j/\l in {0/0, 0.5/1, 1/2} {
        \coordinate (v\k\l) at (\i, \j);
      }
    }
    \draw[thick, black, line join=miter] (v00) -- (v10);
    \draw[nedof, thick, black, line join=miter] (v10) -- (v20);
    \draw[nedof, thick, black, line join=miter] (v20) -- (v21);
    \draw[thick, black, line join=miter] (v21) -- (v22);
    \draw[thick, black, line join=miter] (v22) -- (v12);
    \draw[nedof, thick, black, line join=miter] (v12) -- (v02);
    \draw[nedof, thick, black, line join=miter] (v02) -- (v01);
    \draw[thick, black, line join=miter] (v01) -- (v00);

    \draw[nedof, thick, black, line join=cap] (v01) -- (v11);
    \draw[nedof, thick, black, line join=cap] (v11) -- (v21);
    \draw[nedof, thick, black, line join=cap] (v10) -- (v11);
    \draw[nedof, thick, black, line join=cap] (v11) -- (v12);
    \draw[nedof, thick, black, line join=cap] (v01) -- (v10);
    \draw[nedof, thick, black, line join=cap] (v02) -- (v11);
    \draw[nedof, thick, black, line join=cap] (v11) -- (v20);
    \draw[nedof, thick, black, line join=cap] (v12) -- (v21);

    \coordinate (i10) at ($(v10) + (90:0.1)$);
    \coordinate (i20) at ($(v20) + (135:0.141)$);
    \coordinate (i21) at ($(v21) + (180:0.1)$);
    \coordinate (i12) at ($(v12) + (270:0.1)$);
    \coordinate (i02) at ($(v02) + (315:0.141)$);
    \coordinate (i01) at ($(v01) + (0:0.1)$);
    \path[fill=blue, opacity=0.5]
    \convexpath{i01,i02,i12,i21,i20,i10}{1pt};
  \end{scope}

  \end{tikzpicture}
  \caption{Illustration of which degrees of freedom appear in each numbering for the
    vertex-star patch of \cref{fig:hcurl-patch}, with global Dirichlet
    conditions applied on the top and left of the domain. A numbering for the
    degrees of freedom to be updated (left) is always required; if
    multiplicative updates are selected, degrees of freedom on the
    patch boundary that are not on the global boundary (center) must
    also be numbered; finally, for nonlinear problems, a numbering for
    all degrees of freedom supported on the patch is built (right).\label{fig:patch-numbering}}
\end{figure}
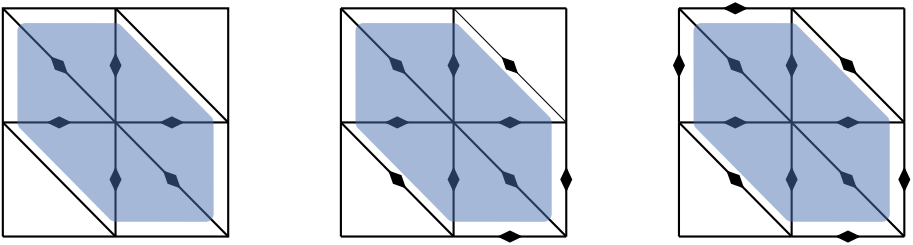

The numberings required on each patch depend on the solution algorithm
employed. In all cases, we build a numbering for the degrees of
freedom that are updated with each patch solve. This excludes all
degrees of freedom on the boundary of the completed patch. In the multiplicative
case, an additional numbering is necessary to describe the degrees of
freedom in the global residual that are updated with each patch solve.
This includes degrees of freedom on the boundary of the completed patch that are
not fixed by global Dirichlet conditions. Finally, in the nonlinear
case, we also require a numbering for the state vector on the patch.
This includes all degrees of freedom in the completed patch, including
those subject to global Dirichlet conditions. The degrees of freedom
involved in each case are illustrated in
\cref{fig:patch-numbering}. Only those numberings
necessary for the solution algorithm employed are constructed. In the
PETSc \texttt{\href{https://www.mcs.anl.gov/petsc/petsc-current/docs/manualpages/DM/PetscDS.html}{\underline{DS}}} implementation, a \texttt{\href{https://www.mcs.anl.gov/petsc/petsc-current/docs/manualpages/PetscSection/PetscSection.html}{\underline{Section}}} numbering object is created for each
patch in the same way that a \texttt{Section} is used for the global
function \citep[\S 10.2]{petsc-user-ref}.

PETSc as a whole, and \dmplex in particular, are well-suited to
parallel computations. The topological mesh is domain-decomposed
across participating parallel processes. With a good decomposition of
the mesh, parallel load balance is obtained and the data structures
have been used with excellent scalability at
$\mathcal{O}(10000\text{--}100000)$ processes
\citep{Hapla2020dmplexio,Parsani2021}. With this domain-decomposition
to hand, assembly of residuals and Jacobians is naturally parallel
across processes.

Globally multiplicative updates require frequent synchronisation
between the parallel subdomains to communicate updated residuals.
As a consequence, in parallel \pcpatch offers either additive or
multiplicative relaxation within a subdomain, but only additive updates
between subdomains. If multiplicative updates within a subdomain are
chosen, as the number of processes is increased the convergence will
approach that of the additive mode. This has the consequence that
additive local updates should be performed if convergence rates independent
of the parallel decomposition are desired.

We require that the mesh is decomposed with enough overlap such that
any patch problem can be constructed entirely locally. This choice avoids
excessive communication when assembling patches.
\dmplex conveniently allows for the specification of the appropriate overlap
when distributing the mesh \citep{lange2015}.
However, this strategy implies a performance tradeoff: larger patches may give better
convergence, but they cause an increase in the sizes of messages
sent over the network due to increased parallel overlap.
Currently, the user must arrange that the mesh decomposition has
enough overlap for the patch specification before the application of
\pcpatch. A future enhancement will use the patch specification to
define the required overlap and automatically expand the mesh overlap
as necessary using the existing overlap distribution algorithm
provided by \dmplex.

\pcpatch offers two mechanisms for the user to configure the space
decomposition. In the basic case, the user specifies the dimension (or codimension) of
entities to iterate over and a patch construction type. In the more advanced case, the user
provides a callback that explicitly enumerates the entities that each
patch contains. For example, the vertex-star space decomposition of
\cref{fig:hcurl-patch} is obtained by specifying a \texttt{star}
construction type around dimension-0 entities (vertices). \pcpatch
presently provides three pre-configured construction types:
\texttt{star}, which gathers entities the topological star of the given
iteration entity; \texttt{vanka}, which gathers entities in the closure
of the star of the given iteration entity; and \texttt{pardecomp},
which creates a single patch per parallel subdomain containing all
locally visible entities. The callback mechanism can implement more complex
patches, such as
line- and plane-smoothers for advection-dominated problems,
or space decompositions induced by a particular mesh structure such as
Alfeld (barycentric) or Powell--Sabin refinement.
This callback
has access to the full \dmplex object and can therefore
construct arbitrary space decompositions, subject to the constraint that
each patch must be local to a single parallel process.

\section{Applications}\label{sec:examples}

\subsection{\hdiv and \hcurl Riesz maps}\label{sec:hdivcurl}
Consider the following problem:
for a bounded Lipschitz domain $\Omega \subset \mathbb{R}^3$, $\alpha > 0$ and $f \in \left(\hdiv\right)^*$, find $u \in \hdiv$ such that
\begin{subequations}\label{eqn:hdiv}
\begin{alignat}{3}
u - \alpha \nabla \nabla \cdot u &= f && \quad \text{ in } \Omega, \\
                  \nabla \cdot u &= 0 && \quad \text{ on } \partial \Omega.
\end{alignat}
\end{subequations}
For $\alpha = 1$,
this is the application of the \hdiv Riesz map.
Similarly, the application of the \hcurl Riesz map entails solving the problem:
for $f \in \left(\hcurl\right)^*$ and $\alpha = 1$, find $u \in \hcurl$ such that
\begin{subequations}\label{eqn:hcurl}
\begin{alignat}{3}
u + \alpha \nabla \times \nabla \times u &= f && \quad \text{ in } \Omega, \\
                         \nabla \times u &= 0 && \quad \text{ on } \partial \Omega.
\end{alignat}
\end{subequations}
These equations often arise as subproblems in the construction of fast
preconditioners for more complex systems involving solution variables in \hdiv
and \hcurl \citep{mardal2011}. However, the operators $I - \alpha \nabla \nabla
\cdot$ and $I + \alpha \nabla \times \nabla \times$ are not elliptic due to the
infinite-dimensional kernels of the divergence and curl operators, and hence
standard relaxation schemes such as Jacobi and Gau\ss--Seidel iteration do not
yield effective multigrid schemes. This can be overcome by employing relaxation
schemes where the space decomposition does capture the kernel, as in~\eqref{eqn:lee2007}. Arnold, Falk, and Winther proved that the star iteration
constructed around vertices does satisfy~\eqref{eqn:lee2007} and hence yields a
mesh-independent and $\alpha$-robust multigrid scheme for both problems
\citep{arnold2000}. The key solver options for expressing the vertex-star relaxation
scheme are given in \cref{lst:vertexstar}.
A complete Firedrake code for solving this problem can be found in \cref{sec:hdivcomplete}.
All calculations for the examples in this section were performed on 28 cores of a
28-core Intel Xeon Gold 5120 CPU @2.20GHz workstation with 192GB of
RAM running Ubuntu 18.04.3.

\begin{listing}
\begin{minted}{python}
      solver_parameters = {
          ...
          "ksp_type": "cg",
          "pc_type": "mg",
          "mg_levels_ksp_type": "richardson",
          "mg_levels_pc_type": "python",
          "mg_levels_pc_python_type": "firedrake.PatchPC",
          "mg_levels_patch_pc_patch_local_type": "additive",
          "mg_levels_ksp_richardson_scale": 1/3,
          "mg_levels_patch_pc_patch_construct_type": "star",
          "mg_levels_patch_pc_patch_construct_dim": 0,
          ...
      }
\end{minted}
\caption{Solver options for \pcpatch to implement a damped additive vertex-star iteration.}
\label{lst:vertexstar}
\end{listing}

\begin{table}[htbp]
  \centering
  \input{code/hdivcurl/table-Hdiv-dim0.tex}
\caption{Number of conjugate gradient iterations to solve the Raviart--Thomas
discretisation of~\eqref{eqn:hdiv} using multigrid and the vertex-star
relaxation.\label{tab:hdiv}}
\end{table}

We present results for the vertex-star relaxation applied to the
lowest-order \hdiv-conforming Raviart--Thomas discretisation of~\eqref{eqn:hdiv}
with varying mesh refinement and $\alpha$ in \cref{tab:hdiv}. The domain and data
were given by $\Omega = (0, 2)^3$ and $f = \left(2yz(1-x^2), 2xz(1-y^2),
2xy(1-z^2)\right)$. The solver employed the conjugate gradient method preconditioned by
a full multigrid cycle, with the relaxation applied additively with a damping of
$1/3$. The
coarse grid had $5 \times 5 \times 5$ elements and the solver was deemed to have
converged when the preconditioned residual norm had decreased by ten
orders of magnitude. As expected from the theory of Arnold et al., the solver
enjoys both mesh-independence and $\alpha$-robustness.

For the \hdiv problem~\eqref{eqn:hdiv}, Arnold et al.~also prove robustness of the edge-star
relaxation. This can be expressed using the option \mintinline{python}{"mg_levels_patch_pc_patch_construct_dim": 1}. Similarly robust results were observed for this space
decomposition, applied additively with a damping of $1/4$ (not shown). However, the absolute
convergence was poorer, with typically 40--47 Krylov iterations required for
convergence instead of 14--16.

\begin{table}[htbp]
\centering
  \input{code/hdivcurl/table-Hcurl-dim0.tex}
\caption{Number of conjugate gradient iterations to solve the N\'ed\'elec
discretisation of~\eqref{eqn:hcurl} using multigrid and the
vertex-star relaxation.\label{tab:hcurl}}
\end{table}

The results for the analogous experiment with vertex-star relaxation applied
to the lowest-order \hcurl-conforming discretisation of~\eqref{eqn:hcurl}
with N\'ed\'elec elements of the first kind are presented in \cref{tab:hcurl}.
The same domain, data, and solver were employed, except that the additive vertex-star
relaxation was applied with a damping of $1/2$.
The solver is again mesh-independent and $\alpha$-robust.
For this problem the edge-star relaxation is not effective (in fact it coincides
with standard Jacobi or Gau\ss--Seidel).

The vertex-star relaxation also yields mesh independent and $\alpha$-robust
results for higher-order discretisations, as well as
for Brezzi--Douglas--Marini elements and N\'ed\'elec elements of the second kind.
\subsection{Nearly incompressible linear elasticity}
The equations of linear elasticity in the isotropic homogeneous case
can be written as: given $\Omega \subset \mathbb{R}^d$ and $f \in \left(\honezerov\right)^*$, find $u \in \honezerov$ such that
\begin{subequations}\label{eqn:ip}
\begin{alignat}{3}
- \mu \nabla \cdot E(u) - \gamma \nabla \nabla \cdot u &= f && \quad \text{ in } \Omega,
\end{alignat}
\end{subequations}
where $E(u) = \frac{1}{2}\left(\nabla u + \nabla u^T\right)$. Here $\mu$ and $\gamma$ are positive real parameters describing the material in question. The difficult case is when the material is nearly incompressible, with $\gamma \to \infty$.
As
guided by the abstract theory, the central task is to
choose a space decomposition $V_h = \sum_i V_i$ that captures the kernel of
the divergence operator. For \honev-conforming Lagrange elements this is a
subtle question that depends on the dimension $d$ and polynomial degree $k$; for
brevity we only consider the case $d = 2$ here.

An important insight is gained by studying the de Rham complex in two dimensions
\begin{equation}
  \label{eqn:derham2d}
  \mathbb{R} \rightarrow \htwo \xrightarrow{\nabla^\perp} \honev \xrightarrow{\nabla \cdot} \ltwo
  \rightarrow 0.
\end{equation}
If $\Omega$ is simply connected then this de Rham complex is exact, i.e.~every
divergence-free function in \honev is the rotated gradient of a function in
\htwo.  The existence of a suitable space decomposition for degree $k$ vector
fields thus hinges on the existence of a local basis for
$C^1(\Omega)$-conforming degree $k+1$ scalar fields. An important result of
\citet{morgan1975} guarantees the existence of a local basis
that is captured by a vertex-star iteration for $k \ge 4$. This vertex-star
iteration can be easily implemented with \pcpatch via the options given
in \cref{lst:vertexstar}.

We consider the continuous Lagrange discretisation of~\eqref{eqn:ip} with
$\Omega = (0, 1)^2$, $f = (1, 1)$, $\mu = 1$ and $\gamma$ varying from $10^0$ to
$10^4$. The degree of the polynomials employed on each cell $k$ was varied from $k =
1, \dots, 6$. The problem was solved with the conjugate gradient method preconditioned
by a full multigrid cycle and vertex-star relaxation with damping of $1/3$. The base mesh was a
$10 \times 10$ grid and five levels of mesh refinement were used for all
problems.

\begin{table}[htbp]
  \centering
  \input{code/elasticity/table-elasticity.tex}
\caption{Number of conjugate gradient iterations to solve the continuous
Lagrange discretisation of~\eqref{eqn:ip}
using full multigrid cycles and additive vertex-star relaxation. For $k \ge 4$
the vertex-star iteration captures the kernel of the divergence operator,
while for $k < 4$ it does not.\label{tab:ip}}
\end{table}

The results are presented in \cref{tab:ip}. The results are strikingly
different for $k < 4$ and $k \ge 4$. For $k < 4$, the solver exhibits
strong $\gamma$-dependence, with the iteration counts blowing up as
$\gamma$ increases. However, for $k \ge 4$, the iteration counts are
$\gamma$-robust. This is exactly as one would expect from the results
of Morgan and Scott, which guarantee that the star iteration captures
the kernel in the sense of~\eqref{eqn:lee2007} for $k \ge 4$. This is
a striking illustration of the sharpness of the abstract theory.

\subsection{Stokes equations}

We consider the incompressible Newtonian Stokes equations: given $\mu > 0$, a bounded
Lipschitz domain $\Omega \subset \mathbb{R}^d$, $f \in \left(\honev\right)^*$
and $g \in \honehalfDv$, find the velocity and pressure $(u, p) \in V \times Q := \honev \times \ltwo$ such that
\begin{subequations}\label{eqn:stokes}
\begin{alignat}{3}
- \nabla \cdot 2\mu E(u) + \nabla p &= f && \quad \text{ in } \Omega, \\
                   \nabla \cdot u &= 0 && \quad \text{ in } \Omega, \\
                          u &= g && \quad \text{ on } \partial \Omega_D, \\
                          \left(-pI + \mu E(u)\right)\cdot n &= 0 && \quad \text{ on } \partial \Omega_N,
\end{alignat}
\end{subequations}
where $n$ is the outward unit normal to $\partial \Omega = \partial \Omega_D \cup \partial \Omega_N$, and $I$ is the $d \times d$ identity matrix.
If $|\partial \Omega_N| = 0$ then the pressure is only defined up to a constant and the pressure trial
space $Q = \ltwozero$ is used instead. The pressure acts as a Lagrange multiplier for enforcing the divergence
constraint.
This equation is a fundamental problem in fluid mechanics and a great many discretisations and solvers
have been proposed for it \citep{turek1999,brandt2011,elman2014}. In this section we discuss the
implementation of
a monolithic multigrid method with Vanka relaxation
\citep{vanka1986}.

Monolithic multigrid methods apply the multigrid philosophy to the entire block-structured
problem (i.e.~solving for velocity and pressure together). It is therefore necessary to
develop appropriate relaxation methods that dampen the error of the coupled problem.
Note that Jacobi or Gau\ss--Seidel would not work for an inf-sup stable discretisation of~\eqref{eqn:stokes}, as the saddle point structure means that there are zero entries on the
main diagonal. One popular strategy for implementing appropriate relaxation methods for
saddle point problems is
Vanka relaxation, where patches are defined by gathering all degrees of freedom connected
to a single degree of freedom of the Lagrange multiplier \citep{maclachlan2011}.

We consider the classical regularised lid-driven cavity benchmark \citep[\S 3.1.3]{elman2014},
discretised using the inf-sup stable CG2-CG1 Taylor--Hood discretisation.
As shown in \cref{fig:vanka-taylor-hood}, the associated Vanka relaxation is defined
by the closure of the star around each vertex (as the pressure degrees of freedom are located
at vertices). The key solver options are given in \cref{lst:vanka-taylor-hood}.
\begin{listing}
\begin{minted}{python}
      solver_parameters = {
          ...
          "ksp_type": "gmres",
          "pc_type": "mg",
          "mg_levels_ksp_type": "chebyshev",
          "mg_levels_ksp_max_it": 2,
          "mg_levels_pc_type": "python",
          "mg_levels_pc_python_type": "firedrake.PatchPC",
          "mg_levels_patch_pc_patch_local_type": "additive",
          "mg_levels_patch_pc_patch_partition_of_unity": False,
          "mg_levels_patch_pc_patch_construct_type": "vanka",
          "mg_levels_patch_pc_patch_construct_dim": 0,
          "mg_levels_patch_pc_patch_exclude_subspaces": "1",
          ...
      }
\end{minted}
\caption{Solver options for \pcpatch to implement Vanka relaxation for a Taylor--Hood discretisation.\label{lst:vanka-taylor-hood}}
\end{listing}
The solver employed GMRES as the outer Krylov solver preconditioned by multigrid
V-cycles, using two iterations of Chebyshev-accelerated Vanka relaxation on each
multigrid level. The pressure nullspace was dealt with by explicitly passing a
basis for the nullspace to the Krylov method (in this case, the vector of
constant pressures). We highlight the option
\mintinline{python}{"mg_levels_patch_pc_patch_exclude_subspaces": "1"}; this
excludes from the patch pressure degrees of freedom (in the subspace indexed by
1) other than that at the vertex around which the patch is built, ensuring that
each patch contains exactly one pressure degree of freedom. Without this, the
patch would include the pressures at other vertices connected by an edge to the
base vertex. The base grid was a uniform 20$\times$20 grid of quadrilaterals.
The solver was deemed to have converged when the Euclidean norm of the residual
had decreased by ten orders of magnitude. The results are shown in \cref{tab:stokes}.
\begin{table}[htbp]
  \centering
  \input{code/stokes/table-stokes.tex}
\caption{Number of GMRES iterations to solve the Taylor--Hood
discretisation of~\eqref{eqn:stokes} using multigrid and Vanka relaxation.\label{tab:stokes}}
\end{table}
The solver enjoys mesh independence and $\nu$-robustness. However, these properties do not
hold for Vanka relaxation applied to the Navier--Stokes equations \citep{turek1999} and a more complex algorithm must be used to achieve
$\nu$-robustness in this case \citep{benzi2006, farrell2018b}.

\subsection{Semilinear Allen--Cahn equation}
As a final example, we consider the semilinear Allen--Cahn equation: given $\Omega \subset \mathbb{R}^d$
and $f \in H^{-1}$, find $u \in H^1_0(\Omega)$ such that
\begin{equation} \label{eqn:allen-cahn}
-\nabla^2 u + u^3 - u = f.
\end{equation}
We consider full approximation scheme (FAS) nonlinear multigrid methods for high-order discretisations of this problem \citep{brandt2011}.
While PETSc has had an implementation of
FAS since 2011, no general nonlinear relaxation methods were
available; users had to implement by hand a custom
nonlinear Gau\ss--Seidel algorithm on a case-by-case basis. This deficiency is remedied by \snespatch, the nonlinear analogue of \pcpatch.

As with any multigrid scheme, the choice of appropriate relaxation is key. When nonlinear problems
are solved with Newton's method on each patch, the residual and Jacobian on the patch must be calculated at each
iteration. If many patches are present this assembly cost can be quite expensive \citep{brabazon2014}.
We therefore compare two different relaxation methods: a nonlinear star iteration\footnote{Since a star
iteration and pointwise Jacobi must assemble over the same cells for each nonlinear iteration, it makes sense to update all
degrees of freedom in the star at the same time.}, and an overlapping Schwarz iteration induced by the
parallel decomposition \citep{dryja1997}. Specifically, in the latter scheme a nonlinear problem is solved independently
on each core, and the updates averaged on the overlap, using Newton's method and LU factorisation as the inner solver. The overlap
is of closure-star type, i.e.~the degrees of freedom on the closure of the star of all entities that
the process owns are solved for.

We consider a continuous Lagrange discretisation of varying degree $p$ of~\eqref{eqn:allen-cahn}. We employ
quadrilateral elements to take advantage of sum factorisation \citep{homolya2017} and tensor-product
Gau\ss--Legendre--Lobatto quadrature rules \citep{karniadakis2005} for efficiency at high-order.

\begin{figure}[htbp]
  \centering
  \resizebox{0.5\textwidth}{!}{\input{images/ngmresfasstar.tikz}}
  \caption{An outline of the NGMRES/FAS/star algorithm for solving~\eqref{eqn:allen-cahn}.\label{fig:ngmresfasstar}}
\end{figure}
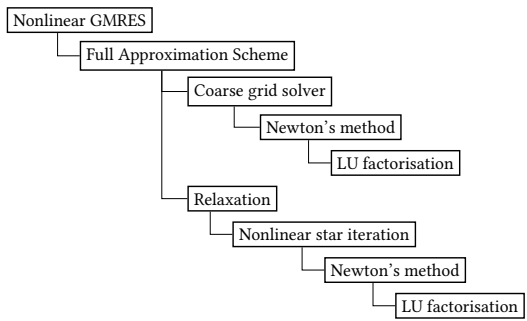

In both solvers, we employ cascadic multigrid cycles \citep{bornemann1996}, using Newton's method
with direct factorisations for both the coarse grid problem and all patch solves. No global matrix
is assembled or solved, only small local or coarse problems. All relaxation updates are performed
with partition of unity weighting. In both cases we use FAS as a nonlinear right preconditioner for
a nonlinear GMRES iteration, analogous to preconditioning linear GMRES using a standard multigrid
scheme \citep{oosterlee2000,brune2015}. For illustration, a diagram of the solver with star relaxation
is given in \cref{fig:ngmresfasstar}.

For reference, we also compare to a standard Newton-Krylov scheme
where the linearised Jacobians are solved with FGMRES, preconditioned
by matrix-free cascadic multigrid cycles with two
Chebyshev-accelerated Jacobi sweeps as relaxation. This latter scheme
does not use \pcpatch; the diagonal of the matrix is assembled
directly. The solves were performed with 4 multigrid refinements of a
uniform 20 $\times$ 20 base grid. The results are shown in
\cref{tab:allen-cahn}.

\begin{table}[htbp]
  \centering
  \input{code/allen-cahn/table-allen-cahn.tex}
\caption{Time in seconds/number of outer nonlinear iterations
for three different solver strategies applied to
discretisations of varying degree $p$ of~\eqref{eqn:allen-cahn} using multigrid and Vanka relaxation.\label{tab:allen-cahn}}
\end{table}

For $p \le 5$, FAS with large patches is more efficient than FAS with small patches;
for larger $p$, FAS with star relaxation is substantially more efficient. However,
in all cases the schemes based on FAS are not competitive with Newton--Krylov for this
problem.
We expect that the code would be even faster if nonlinear
star iteration were used as a nonlinear preconditioner for Newton's method
\citep{cai2002}.

\section{Conclusion}\label{sec:conclusion}

We have presented \pcpatch, a multigrid relaxation framework whose
software abstractions are motivated by the abstract subspace
corrections framework of \citet{xu1992}. By using a callback, rather
than algebraic, interface \pcpatch supports nonlinear and matrix-free
relaxation in the same generic way. By maintaining topological
information in multigrid hierarchies, and utilising a simple
topological query language, it offers a flexible and extensible
interface to a broad range of optimal relaxation methods.
We have demonstrated its application for the easy development of
efficient multigrid solvers for both linear and non-linear problems.

\section*{Code availability}
For reproducibility, we cite archives of the exact software versions used to
produce the results in this paper. All major Firedrake components as well as
the code used to obtain the shown iteration counts and runtimes have been
archived on Zenodo~\citep{pcpatch-code}. An installation of Firedrake with
components matching those used to produce the results in this paper can by
obtained following the instructions at
\url{https://www.firedrakeproject.org/download.html}.

\begin{acks}
This work is
supported by the \grantsponsor{EP/R029423/1}{Engineering and Physical
Sciences Research Council}{https://gow.epsrc.ukri.org} under grants
\grantnum{EP/R029423/1}{EP/R029423/1}, \grantnum{EP/V001493/1}{EP/V001493/1}, and \grantnum{EP/L015803/1}{EP/L015803/1}. LM also acknowledges support
from the \grantsponsor{EP/N032861/1}{UK Fluids
  Network}{https://gow.epsrc.ukri.org/NGBOViewGrant.aspx?GrantRef=EP/N032861/1}
[EPSRC grant number \grantnum{EP/N032861/1}{EP/N032861/1}] for funding a visit
to Oxford. MGK acknowledges partial funding for this work from \grantsponsor{DE-AC02-06CH11357}{U.S. Department of Energy}{https://app.dimensions.ai/details/grant/grant.7824474} under grant \grantnum{DE-AC02-06CH11357}{DE-AC02-06CH11357}.
\end{acks}

\appendix
\section{Complete code listing for the \hdiv Riesz map}\label{sec:hdivcomplete}
\inputminted[mathescape=true]{python}{hdiv.py}

\ifarxiv
\input{paper.bbl}

\else
\bibliographystyle{ACM-Reference-Format}
\bibliography{paper}
\fi
\end{document}

%% file: code/hdivcurl/table-Hdiv-dim0.tex
\begin{tabular}{cc|ccccc}
\toprule
\# refinements & \# degrees of freedom & \multicolumn{5}{c}{$\alpha$} \\
 && $10^0$ & $10^1$ & $10^2$ & $10^3$ & $10^4$ \\
\midrule
1 & $1.26 \times 10^{4}$ & 14 & 14 & 14 & 14 & 14\\
2 & $9.84 \times 10^{4}$ & 15 & 15 & 15 & 15 & 15\\
3 & $7.78 \times 10^{5}$ & 15 & 16 & 16 & 16 & 16\\
4 & $6.18 \times 10^{6}$ & 16 & 16 & 16 & 16 & 16\\
\bottomrule
\end{tabular}

%% file: code/hdivcurl/table-Hcurl-dim0.tex
\begin{tabular}{cc|ccccc}
\toprule
\# refinements & \# degrees of freedom & \multicolumn{5}{c}{$\alpha$} \\
 && $10^0$ & $10^1$ & $10^2$ & $10^3$ & $10^4$ \\
\midrule
1 & $7.93 \times 10^{3}$ & 18 & 19 & 19 & 19 & 19\\
2 & $5.97 \times 10^{4}$ & 19 & 19 & 19 & 19 & 19\\
3 & $4.63 \times 10^{5}$ & 20 & 20 & 20 & 20 & 20\\
4 & $3.64 \times 10^{6}$ & 20 & 20 & 20 & 20 & 20\\
\bottomrule
\end{tabular}

%% file: code/elasticity/table-elasticity.tex
\begin{tabular}{cc|ccccc}
\toprule
$k$ & \# degrees of freedom & \multicolumn{5}{c}{$\gamma$} \\
 && $10^0$ & $10^1$ & $10^2$ & $10^3$ & $10^4$ \\
\midrule
1 & $2.06 \times 10^{5}$ & 32 & 68 & 189 & 541 & $>1000$\\
2 & $8.22 \times 10^{5}$ & 15 & 28 & 74 & 256 & $>1000$\\
3 & $1.85 \times 10^{6}$ & 12 & 18 & 37 & 102 & 310\\
\midrule
4 & $3.28 \times 10^{6}$ & 10 & 13 & 16 & 16 & 16\\
5 & $5.13 \times 10^{6}$ & 9 & 11 & 14 & 14 & 13\\
6 & $7.38 \times 10^{6}$ & 9 & 11 & 12 & 12 & 11\\
\bottomrule
\end{tabular}

%% file: code/stokes/table-stokes.tex
\begin{tabular}{cc|ccccc}
\toprule
\# refinements & \# degrees of freedom & \multicolumn{5}{c}{$\nu$} \\
 && $10^0$ & $10^1$ & $10^2$ & $10^3$ & $10^4$ \\
\midrule
1 & $1.48 \times 10^{4}$ & 14 & 14 & 14 & 14 & 14\\
2 & $5.84 \times 10^{4}$ & 14 & 14 & 14 & 14 & 14\\
3 & $2.32 \times 10^{5}$ & 14 & 14 & 14 & 14 & 14\\
4 & $9.25 \times 10^{5}$ & 14 & 14 & 14 & 14 & 14\\
\bottomrule
\end{tabular}

%% file: images/ngmresfasstar.tikz
\begin{tikzpicture}[%
  every node/.style={draw=black, thick, anchor=west},
  grow via three points={one child at (0.0,-0.7) and
  two children at (0.0,-0.7) and (0.0,-1.4)},
  edge from parent path={(\tikzparentnode.210) |- (\tikzchildnode.west)}]
  \node {Nonlinear GMRES}
    child { node {Full Approximation Scheme}
      child { node {Coarse grid solver}
        child { node {Newton's method}
          child { node {LU factorisation}}
        }
      }
      child [missing] {}
      child [missing] {}
      child { node {Relaxation}
        child { node {Nonlinear star iteration}
          child { node {Newton's method}
            child { node {LU factorisation}}
          }
        }
      }
    };
\end{tikzpicture}

%% file: code/allen-cahn/table-allen-cahn.tex
\begin{tabular}{cc|ccc}
\toprule
$p$ & \# degrees of freedom & \multicolumn{3}{c}{Solver type} \\
 && {\footnotesize NGMRES/FAS/star} & {\footnotesize NGMRES/FAS/pardecomp} & {\footnotesize Newton/FGMRES/MG/Jacobi} \\
\midrule
3 & $9.24 \times 10^{5}$ & 6.5/5 & 4.0/3 & 1.5/2\\
4 & $1.64 \times 10^{6}$ & 8.6/4 & 6.2/2 & 2.2/2\\
5 & $2.56 \times 10^{6}$ & 11.6/4 & 8.7/1 & 4.0/2\\
6 & $3.69 \times 10^{6}$ & 13.9/3 & 17.1/1 & 6.3/2\\
7 & $5.02 \times 10^{6}$ & 19.5/3 & 27.5/1 & 9.6/2\\
8 & $6.56 \times 10^{6}$ & 21.9/2 & 45.0/1 & 12.6/2\\
9 & $8.30 \times 10^{6}$ & 32.2/2 & 71.3/1 & 18.3/2\\
10 & $1.02 \times 10^{7}$ & 42.5/2 & 115.7/1 & 24.5/2\\
\bottomrule
\end{tabular}